\documentclass[12pt]{article}

\usepackage{amssymb,amsfonts,amsthm}
\usepackage{graphicx}
\usepackage{subfigure}
\usepackage{amssymb}
\usepackage{amsthm}
\usepackage{amsmath}
\usepackage{bm}             
\usepackage[mathscr]{eucal}

\usepackage{cancel}

\usepackage{psfrag}

\usepackage{multirow}

\usepackage{wasysym}

\usepackage{color}

\def\pmb#1{\setbox0=\hbox{$#1$}%
  \kern-.025em\copy0\kern-\wd0
  \kern.05em\copy0\kern-\wd0
  \kern-.025em\raise.0433em\box0}
\def\pmbs#1{\setbox0=\hbox{$\scriptstyle #1$}%
  \kern-.0175em\copy0\kern-\wd0
  \kern.035em\copy0\kern-\wd0
  \kern-.0175em\raise.0303em\box0}

\def\be{\begin{equation}}
\def\ee{\end{equation}}
\def\bea{\begin{eqnarray}}
\def\eea{\end{eqnarray}}

\def\hsp5{\hspace{5mm}}
\def\case#1/#2{\textstyle\frac{#1}{#2}}

\theoremstyle{plain}

\theoremstyle{remark}

\newtheorem*{remark}{Remark}

\newcommand{\sfrac}[2]{{\textstyle{#1\over#2}}}

\title{\sc Global cosmological dynamics for the scalar field representation of the modified Chaplygin gas}

\begin{document}

\author{
Claes Uggla\thanks{Electronic
address:{\tt claes.uggla@kau.se}}\\
{\small\em Department of Physics, University of Karlstad,}\\
{\small\em S-65188 Karlstad, Sweden}}

\date{}
\maketitle

\begin{abstract}

In this paper we investigate the global dynamics for the minimally coupled
scalar field representation of the modified Chaplygin gas in the context of
flat Friedmann-Lema\^{i}tre-Robertson-Walker cosmology. The tool for doing
this is a new set of bounded variables that lead to a regular dynamical
system. It is shown that the exact modified Chaplygin gas perfect fluid
solution appears as a straight line in the associated phase plane. It is also
shown that no other solutions stay close to this solution during their entire
temporal evolution, but that there exists an open subset of solutions that
stay arbitrarily close during an intermediate time interval, and into the
future in the case the scalar field potential exhibits a global minimum.

\end{abstract}

\centerline{\bigskip\noindent PACS numbers: 04.20.-q, 98.80.-k,
98.80.Bp, 98.80.Jk}

\section{Introduction}

The total matter content in the universe, now and in the distant past, is a
mystery. As a consequence cosmological models abound, both as regards fields
and the dynamical laws they obey. This is reflected in a plethora of
inflationary models and models for dark matter and energy, in the context of
general relativity and in modified gravity theories. This entails
possibilities for new physical insights, but it also causes explanatory
problems; the latter exemplified by e.g. fine-tuning problems of potentials
(see e.g.~\cite{ijj13}) and initial data.

One attempt to fit theory with observations is to phenomenologically match
the observed expansion history by imposing conditions on e.g. the time
development of the cosmological scale factor, or by imposing a relationship
between pressure and energy density, and then use such a relationship to
produce a field description, as discussed in
e.g.~\cite{sahsta06}--\cite{bametal12} and references therein. The arguably
simplest field theoretic description is a minimally coupled scalar field
within the context of flat Friedmann-Lema\^{i}tre-Robertson-Walker (FLRW)
cosmology, see e.g.~\cite{ratpeb88}--\cite{capste13}, and references therein.
A field theoretic description, however, leads to that the original
relationship only is obeyed by particular solutions to the field problem it
has generated, as pointed out in e.g.~\cite{janetal93}--\cite{goretal05}.
This is due to that a field theoretic description adds degrees of freedom,
e.g. a single minimally coupled scalar field adds one degree of freedom to a
perfect fluid description with a barotropic equation of state. This then
leads to the issue how the global solution space of the field theoretic
description is related to that of the generating relationship.

Although some work on the issue of initial data not corresponding to the
generating conditions has been done, as exemplified by the numerical examples
in~\cite{peretal04} and the investigation in~\cite{goretal05} for the
Chaplygin gas, as far as the author knows there seems to be no global
dynamical systems investigations of this issue. In this paper, which is the
first in a series that will address fine-tuning problems with global
dynamical systems methods, we will, as a specific example, consider flat FLRW
cosmology and the minimally coupled scalar field description of the modified
Chaplygin gas. The modified Chaplygin gas is characterized by an equation of
state (see e.g.~\cite{debetal04}--\cite{benetal06})

\begin{equation}\label{Chapeqstate}
p = (\gamma - 1)\rho - M\rho^{-\mu},
\end{equation}
where $p$ is the pressure of the fluid, $\rho$ its energy density and $M$,
$\mu$ and $\gamma$ are free parameters. For simplicity, we will here restrict
their range to $M>0$, $\mu> 0$, $1\leq \gamma <2$. The value $\gamma=1$ leads
to the so-called generalized Chaplygin gas while $\gamma=\mu=1$ corresponds
to the original Chaplygin gas. This equation of state can be used to
construct the scalar field potential for a scalar field $\phi$, see
e.g.~\cite{kametal01,goretal03,debetal04,cos08}, which can be written as
\begin{equation}\label{Vpot}
V = \frac{V_0}{2}\left[(2-\gamma)\cosh^{\frac{2}{1+\mu}}\tilde{\phi} + \gamma\cosh^{\frac{-2\mu}{1+\mu}}\tilde{\phi}\right]
= \frac{V_0}{2}\cosh^{\frac{2}{1+\mu}}\tilde{\phi}\left[2 - \gamma\tanh^{2}\tilde{\phi}\right],
\end{equation}
where $V_0:= (M/\gamma)^{\frac{1}{1+\mu}}$, $\tilde{\phi}:=\frac12
(1+\mu)\sqrt{3\gamma}(\phi - \phi_0)$.

In this paper we will introduce new bounded variables that in the flat FLRW
case with the above minimally coupled scalar field lead to a 2-dimensional
regularized dynamical system, with $\mu$ and $\gamma$ as parameters. This
will yield a global picture of the scalar field dynamics, and provide a
context for the generating modified Chaplygin gas perfect fluid solution,
which, for a given $\mu$ and $\gamma$, appears as a straight line in our
regularized 2-dimensional dynamical system. Moreover, the scalar field
potential~\eqref{Vpot} yields three classes of different behavior,
characterized by conditions on $\mu$ and $\gamma$ associated with a
bifurcation for the dynamical system; notably the original Chaplygin gas
appears as a member of the class of models associated with the bifurcation.

The outline of the paper is as follows. In the next section we derive, in a
step by step manner, our regularized dynamical system on a bounded state
space, where the boundaries are included. In section~\ref{dynsysanalysis} we
perform a both local and global dynamical systems analysis which leads to a
complete understanding of the solution space. In particular the generating
perfect fluid is identified as an invariant subset corresponding to a
straight line in the phase plane, and this subset is subsequently
contextualized by the local and global phase plane results. The paper is
concluded with a discussion in section~\ref{disc}.

\section{Dynamical systems formulation}\label{dynsysappr}

The field equations for a minimally coupled scalar field $\phi(t)$ with
potential $V(\phi)$ for flat FLRW cosmology are given by
\begin{subequations}\label{Hphieq}
\begin{align}
3H^2 &= \sfrac12 \dot{\phi}^2 + V =: \rho_\phi,\label{Gauss1}\\
\dot{H} &= -H^2 - \sfrac{1}{3}(\dot{\phi}^2 - V) = -\sfrac{1}{2} \dot{\phi}^2,\label{Ray1}\\
0 &= \ddot{\phi} + 3H\dot{\phi} + V_{\phi}.
\end{align}
\end{subequations}
Here an overdot signifies the synchronous time, $t$, derivative; $V_{\phi}:=
dV/d\phi$, and where we have used units to set $c=1=8\pi G$, where $c$ is the
speed of light and $G$ is the gravitational constant. The Hubble variable $H$
is given by $H = \dot{a}/a$, where $a(t)$ is the cosmological scale factor;
throughout we will assume an expanding Universe, i.e. $H>0$. It follows
from~\eqref{Ray1} that the deceleration parameter, $q$, which is defined via
$\dot{H} = -(1 + q)H^2$, is given by
\begin{equation}\label{qscalar}
q = - 1 + \frac{1}{2}\left(\frac{\dot{\phi}}{H}\right)^2.
\end{equation}

A commonly used formulation when $V \geq 0$ is obtained by
`Hubble-normalization' which is used extensively in cosmology (see
e.g.~\cite{copetal06,col03} for FLRW scalar field cosmology, and
e.g.~\cite{col03,waiell97} for anisotropic spatially homogeneous cosmology):
\begin{equation}\label{xydef}
x:= \frac{\dot{\phi}}{\sqrt{6}\,H}\qquad y:= \frac{\sqrt{V}}{\sqrt{3}\,H},
\end{equation}
and a new time variable ${\tau}$, defined by $dt=H^{-1}d{\tau}$, which means
that ${\tau}= \ln(a/a_0)$ ($a_0=a(t_0)$, where $t_0$ is some convenient
reference time), where $\tau$ sometimes is referred to as $N$, the number of
e-folds from the reference time $t_0$. This leads to:
\begin{subequations}\label{xy}
\begin{align}
x^\prime &= -(3x - \sqrt{\sfrac{3}{2}}\,\lambda)(1-x^2),\\
y^\prime &= (3x - \sqrt{\sfrac{3}{2}}\,\lambda)xy,\\
1 &= x^2 + y^2,\label{HGauss}
\end{align}
\end{subequations}
where a $^\prime$ denotes differentiation with respect to ${\tau}$. The
quantity $\lambda$ is defined by
\begin{equation}
\lambda = - \frac{V_{\phi}}{V},
\end{equation}
and is a function of $\phi$ except when $V = c_1^2\exp(c_2\phi)$, where $c_1$
and $c_2$ are constants, since then $\lambda = -c_2$, which leads to a
1-dimensional problem for $x$. In general, since $\lambda=\lambda(\phi)$, one
has to add the equation
\begin{equation}
\phi^\prime = \sqrt{6} x
\end{equation}
to the system~\eqref{xy} to obtain a closed constrained system. In the above
equations, the variable $y$ can be replaced by a variable $\Omega_V = y^2 =
1- x^2$ which can be globally solved for to yield a 2-dimensional
unconstrained system for $x$ and $\phi$:
\begin{subequations}\label{xphi}
\begin{align}
x^\prime &= -(3x - \sqrt{\sfrac{3}{2}}\,\lambda)(1-x^2),\\
\phi^\prime &= \sqrt{6}\, x.
\end{align}
\end{subequations}
Although $x$ is bounded, $\phi$ is not. Furthermore, $\lambda(\phi)$ need not
be bounded either. In the special case that $\lambda(\phi)$ is bounded for
all $\phi$, one can introduce a new variable $Y(\phi)$ to obtain a new system
\begin{subequations}\label{xY}
\begin{align}
x^\prime &= -(3x - \sqrt{\sfrac{3}{2}}\,\lambda)(1-x^2),\\
Y^\prime &= \sqrt{6}\, \frac{dY}{d\phi}\, x,
\end{align}
\end{subequations}
where, with some abuse of notation $\lambda = \lambda(Y)$ and
$\frac{dY}{d\phi} = \frac{dY}{d\phi}(Y)$. It turns out that it is possible to
choose a bounded variable $Y$ for a number of scalar field potentials so that
the equations~\eqref{xY} become regular on a relatively compact state space,
whose boundary can be included in the dynamical systems analysis, which
allows one to obtain a global picture of the dynamics.

It follows from~\eqref{xydef}, \eqref{HGauss} and~\eqref{xY} that the
variable $x\in (-1,1)$ can be extended to include the boundaries $x=\pm 1$ so
that $x \in [-1,1]$. The invariant boundary subsets $x=\pm 1$, correspond to
$y=0=V$, and are hence associated with the massless scalar field problem, and
we will therefore refer to them as the massless scalar field boundary
subsets, and denote them by ${\cal M_\pm}$. Note that
\begin{equation}
w:= \frac{p_\phi}{\rho_\phi} = \frac{\sfrac12 \dot{\phi}^2 - V(\phi)}{\sfrac12 \dot{\phi}^2 + V(\phi)} = 2x^2 - 1,\qquad q = - 1 + 3x^2,
\end{equation}
and that acceleration ($q<0$) hence occurs if $x^2<\frac13$, which
corresponds to $w<-\frac13$, while $q=2$, $w=1$ on ${\cal M_\pm}$.
Furthermore, it follows from~\eqref{Gauss1} and~\eqref{xydef} that
\begin{equation}
3H^2 = \frac{V(Y)}{1-x^2} =: Z(Y,x),
\end{equation}
while~\eqref{Ray1} gives
\begin{equation}\label{mon1}
Z^\prime = -6x^2\,Z,
\end{equation}
and hence $Z$ is monotonically decreasing towards the future, except when
$x=0$ belongs to an invariant subset. Next we turn to our dynamical systems
formulation of the modified Chaplygin gas scalar field problem.

\begin{figure}[ht]
\centering
        \includegraphics[height=0.40\textwidth]{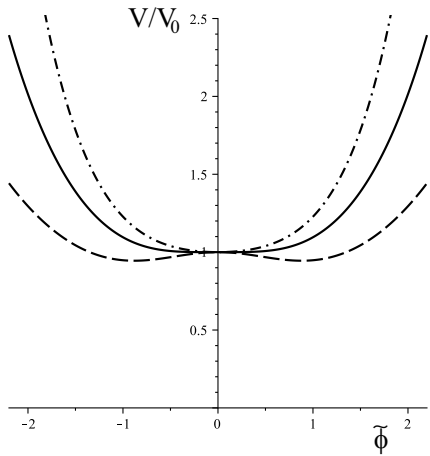}
        \caption{Representatives of the three types of potentials characterized by $\beta>0$, with $\gamma=\frac43,\,\mu=\frac14$ (dashed dotted line);
        $\beta=0$, with $\gamma=1=\mu$ (solid line); $\beta<0$, with $\gamma=1,\,\mu=2$ (dashed line), where
        $\beta= \frac{1}{1+\mu} - \frac{\gamma}{2}$.}
    \label{fig:potentials}
\end{figure}

Recall that the scalar field potential associated with the modified Chaplygin
gas given in~\eqref{Vpot} could be written as
\begin{equation}
V = \frac{V_0}{2}\cosh^{\frac{2}{1+\mu}}\tilde{\phi}\left[2 - \gamma\tanh^{2}\tilde{\phi}\right].
\end{equation}
This potential is invariant under the change $\tilde{\phi} \rightarrow -
\tilde{\phi}$, which will give rise to a discrete symmetry in the dynamical
systems treatment. It is instructive to Taylor expand the potential at
$\tilde{\phi}=0$:
\begin{subequations}
\begin{equation}
V = V_0\!\left[1 + \beta\tilde{\phi}{}^2 + \sfrac18\left((2-\gamma)\gamma + 4\beta(\beta - \sfrac13)\right)\!\tilde{\phi}{}^4 + \dots \right],
\end{equation}
where
\begin{equation}
\beta := \frac{1}{1+\mu} - \frac{\gamma}{2} = \frac{2-\gamma(1+\mu)}{2(1+\mu)}.
\end{equation}
\end{subequations}
The potential has a global minimum at $\tilde{\phi}=0$ if $\beta\geq 0$,
although the minimum is quite `flat' if $\beta=0$ since then
$d^2V/d\tilde{\phi}{}^2|_{\tilde{\phi}=0}=0$, while $\tilde{\phi}=0$ is local
maximum surrounded by two identical minima if $\beta<0$; see
Figure~\ref{fig:potentials}. As can be expected, this will lead to a
bifurcation at $\beta=0$ in the dynamical systems analysis below.

To obtain a regular dynamical system on a relatively compact state space we
introduce the variable\footnote{A variable similar to $Y$ can be used for a
whole range of scalar field problems, although it is not always an optimal
choice; it is chosen here partly because it gives a simple description of the
modified Chaplygin perfect fluid solutions as straight lines.}
\begin{equation}
Y := \tanh\tilde{\phi},
\end{equation}
which leads to that\footnote{As follows from the mathematical properties of
the system~\eqref{xy1}, the range of $\mu$ can be extended from $\mu>0$ to
$\mu>-1$, without any qualitative dynamical changes taking place from a
mathematical point of view. However, at $\mu=-1$ a bifurcation takes place,
which corresponds to that the equation of state of the generating solution
becomes linear.}
\begin{subequations}\label{xy1}
\begin{align}
x^\prime &= -(3x - \sqrt{\sfrac{3}{2}}\,\lambda)(1-x^2),\\
Y^\prime &= \sfrac32(1+\mu)\sqrt{2\gamma}(1-Y^2)x,
\end{align}
where
\begin{equation}
\lambda = -\sqrt{3\gamma}\,Y\left(\frac{2-\gamma - \gamma\mu(1-Y^2)}{2-\gamma Y^2}\right),
\end{equation}
\end{subequations}

Since $\lambda(Y)$ is a differentiable function for $Y\in[-1,1]$, we can
extend the state space to not just including the boundaries $x^2=1$ but also
$Y^2=1$. This leads to a state space ${\bf S}$ for which $x\in (-1,1)$, $Y\in
(-1,1)$, which when extended to include its boundary yields $\bar{\bf S}$:
$x\in [-1,1]$, $Y\in [-1,1]$. In addition to the massless scalar field
boundary subsets ${\cal M}_\pm$, we therefore also have the invariant
boundary subsets $Y=\pm 1$, for which $\lambda(Y=1) = - \lambda(Y=-1)$ is a
constant, and hence the equation for $x$ is the same as that for a single
exponential field potential with $V = c_1^2\exp(-\lambda(Y=\pm1))$. We will
therefore refer to the invariant $Y=\pm 1$ boundary subsets as the
exponential term subsets, and will denote them by ${\cal E}_\pm$.
Furthermore, the system~\eqref{xy1} on $\bar{\bf S}$ admits a discrete
symmetry; it is invariant under the transformation $(x,Y) \rightarrow
-(x,Y)$, a property that is a consequence of the discrete symmetry of the
potential $V$.

In addition to the boundary subsets, the system~\eqref{xy1} admits another
identifiable 1-dimensional subset, namely the interior subset that is
associated with the perfect fluid solution that generated the scalar field
potential. This subset can conveniently be found by using that $w=2x^2-1$ in
general and that $w = w(\phi(Y))$ for the perfect fluid solution, where
$w(\phi(Y))$ is obtained easily from the expressions in
e.g.~\cite{debetal04,cos08}. By identifying the two expressions for $w$, it
then follows straightforwardly that the perfect fluid solution satisfies
\begin{equation}
(\sqrt{2}\,x + \sqrt{\gamma}\,Y)(\sqrt{2}\,x - \sqrt{\gamma}\,Y) = 0,
\end{equation}
and therefore one of these factors must be zero. Taking the combination
$\sqrt{2}\,x^\prime + \sqrt{\gamma}\,Y^\prime$ gives
\begin{equation}\label{pfinv}
(\sqrt{2}\,x + \sqrt{\gamma}\,Y)^\prime = \frac32\!\left(\sqrt{2}\,x +
\sqrt{\gamma}\,Y\right)\!
\left(\sqrt{2}\,x(\sqrt{2}\,x - \sqrt{\gamma}\,Y) - \frac{(2-\sqrt{2\gamma}\,xY)R}{2-\gamma Y^2}\right),
\end{equation}
where
\begin{equation}
R:= 2 - \gamma - \gamma\mu(1-Y^2),
\end{equation}
and hence $\sqrt{2}\,x + \sqrt{\gamma}\,Y=0$ is an invariant subset (which,
by inspection, $\sqrt{2}\,x - \sqrt{\gamma}\,Y=0$ turns out not to be)
describing the perfect fluid solution, which, due to the discrete symmetry,
is represented by two equivalent solution trajectories. We will refer to the
invariant subset $\sqrt{2}x + \sqrt{\gamma}\,Y =0$ as the `perfect fluid
subset' and will denote it as ${\cal PF}$, and the two perfect fluid
trajectories as ${\cal PF}_\pm$, where the sign refers to the sign of $Y$.

The system~\eqref{xy1} admits a number of fixed points:
\begin{subequations}\label{fixedp}
\begin{align}
\mathrm{dS}_0:\quad (x,Y) &= (0,0),\\
\mathrm{dS}_\pm:\quad (x,Y) &= (0,\pm Y_0) \quad Y_0:= \sqrt{1 - \frac{2-\gamma}{\gamma\mu}}, \quad \text{if} \quad \mu > \frac{2-\gamma}{\gamma},\\
\mathrm{M}_{x=\pm 1}^{Y=\epsilon}:\quad (x,Y) &= (\pm 1,\epsilon),\quad \text{where}\quad \epsilon = 1 \quad \text{or} \quad \epsilon =-1,\\
\mathrm{PF}_\pm:\quad (x,Y) &= (\mp \sqrt{\gamma/2},\pm 1) .
\end{align}
\end{subequations}
The fixed points $\mathrm{dS}_0$ and $\mathrm{dS}_\pm$ correspond to $w=-1$
and therefore represent de Sitter states. Note that the fixed points
$\mathrm{dS}_\pm$ only exist when $\mu > \frac{2-\gamma}{\gamma}$, i.e., when
$\beta<0$, which corresponds to the situation when the scalar field potential
$V$ has two minima. The four fixed points $\mathrm{M}_{x=\pm 1}^{Y=\epsilon}$
at the corners of the phase plane have $w=1$ and hence correspond to massless
scalar field states. Finally, the $\mathrm{PF}_\pm$ fixed points are the
origin ($\alpha$ limits) of the perfect fluid solution trajectories ${\cal
PF}_\pm$.\footnote{Note that the fixed points $\mathrm{PF}_\pm$ are
associated with the perfect fluid solution that generates the exponential
scalar field potential which is associated with the ${\cal E}_\pm$ subsets,
i.e., again the generating perfect fluid solution constitutes an invariant
set of codimension one compared to the scalar field state space it
generates.} At these fixed points $w=\gamma-1$, which is the limit of the
perfect fluid equation of state towards the initial singularity when $\rho
\rightarrow \infty$.

The Chaplygin gas case play a historical special role and it is of interest
to write down the equations explicitly for this case, including its fixed
points. In this case $p = - M\rho^{-1}$, and hence $\gamma=\mu=1$, which
leads to the scalar field potential~\cite{kametal01}
\begin{equation}
V = \frac{V_0}{2}\cosh\tilde{\phi}\left[2 - \tanh^{2}\tilde{\phi}\right] =
\frac{V_0}{2}\left[\cosh\tilde{\phi} + \cosh^{-1}\tilde{\phi}\right],
\end{equation}
where $\tilde{\phi}:=\sqrt{3}(\phi - \phi_0)$. As a consequence
\begin{subequations}\label{chapdyn}
\begin{align}
x^\prime &= -(3x - \sqrt{\sfrac{3}{2}}\,\lambda)(1-x^2),\\
Y^\prime &= 3\sqrt{2}(1-Y^2)\,x,\\
\lambda &= -\sqrt{3}\left(\frac{Y^3}{2-Y^2}\right).
\end{align}
\end{subequations}
The ${\cal PF}$ subset is characterized by $\sqrt{2}x + Y = 0$, while the
seven fixed points the dynamical system~\eqref{chapdyn} admits are given by
\begin{subequations}
\begin{align}
\mathrm{dS}_0:\quad (x,Y) &= (0,0),\\
\mathrm{M}_{x=\pm 1}^{Y=\epsilon}:\quad (x,Y) &= (\pm
1,\epsilon),\quad\text{where
}\quad \epsilon =1 \quad \text{or}\quad \epsilon =-1,\\
\mathrm{D}_\pm:\quad (x,Y) &= (\mp 1/\sqrt{2},\pm 1) .
\end{align}
\end{subequations}
The fixed points $\mathrm{PF}_\pm$ have here been denoted by $\mathrm{D}_\pm$
since $w=0$ in this case, and thus $\mathrm{D}_\pm$ represents a `dust'
state.

\section{Dynamical systems analysis}\label{dynsysanalysis}

\subsection{Local fixed point analysis}

The eigenvalues of the various fixed points of~\eqref{xy1} are given by
\begin{subequations}\label{fixedpanalysis}
\begin{alignat}{2}
\mathrm{dS}_0:& \quad -\sfrac32\left[2 - \gamma(1+\mu)\right], &\qquad\quad& -\sfrac32\gamma(1+\mu),\\
\mathrm{dS}_\pm:&\quad -\sfrac32\left[1 \pm \sqrt{9-4\gamma(1+\mu)}\right], &\qquad\quad& \text{if}\quad \mu > \frac{2-\gamma}{\gamma},\\
\mathrm{M}_{x=\mp 1}^{Y=\pm 1}:& \quad 3\sqrt{2\gamma}\,(1+\mu), &\qquad\quad& 3(2-\sqrt{2\gamma}),\\
\mathrm{M}_{x=\pm 1}^{Y=\pm 1}:& \quad -3\sqrt{2\gamma}\,(1+\mu), &\qquad\quad& 3(2+\sqrt{2\gamma}),\\
\mathrm{PF}_\pm:& \quad 3\gamma(1+\mu),&\quad& - \sfrac32(2-\gamma) .
\end{alignat}
\end{subequations}

As expected there is a bifurcation associated with the $\mathrm{PF}_\pm$
fixed points when $\gamma=2$, since $\gamma=2$ leads to that the perfect
fluid solution asymptotically towards the past behave like a stiff fluid,
which is equivalent to a massless scalar field. As a consequence, the
$\mathrm{PF}_\pm$ fixed points pass through the $\mathrm{M}_{x=\mp 1}^{Y=\pm
1}$ points at this value for $\gamma$, which leads to a bifurcation. For
simplicity we have therefore assumed that $\gamma<2$, but we will return to
the case $\gamma=2$ in the concluding discussion.

We also have a bifurcation at $2 - \gamma(1+\mu)=0$ ($\beta=0$), which is
when the $\mathrm{dS}_\pm$ fixed points merge with the $\mathrm{dS}_0$ fixed
point. When $2 - \gamma(1+\mu)<0$ ($\beta<0$) the fixed points
$\mathrm{dS}_\pm$ exist and are hyperbolic sinks, while $\mathrm{dS}_0$ is a
hyperbolic saddle. In this case only two solutions end at $\mathrm{dS}_0$
(has $\mathrm{dS}_0$ as their $\omega$ limit) and these are the perfect fluid
trajectories ${\cal PF}_\pm$, since the invariant subset ${\cal PF}$ can be
identified as the stable manifold of $\mathrm{dS}_0$. When $2 -
\gamma(1+\mu)>0$ ($\beta>0$) the fixed point $\mathrm{dS}_0$ becomes a
hyperbolic sink, however, when $2 - \gamma(1+\mu)=0$ $\mathrm{dS}_0$ is a
non-hyperbolic sink (as will be shown below in the global analysis). Note
that the Chaplygin case belongs to this case. The fixed points
$\mathrm{M}_{x=\mp 1}^{Y=\pm 1}$ are hyperbolic sources while
$\mathrm{M}_{x=\pm 1}^{Y=\pm 1}$ are hyperbolic saddles, with eigenvectors
along the boundaries, so there are no solutions originating from these fixed
points into ${\bf S}$. Finally, $\mathrm{PF}_\pm$ are hyperbolic saddles
where each fixed point gives rise to a single solution entering the state
space ${\bf S}$, the perfect fluid solution ${\cal PF}_\pm$ (thus being the
unstable manifold of $\mathrm{PF}_\pm$).

\subsection{Global dynamical systems analysis}

It follows from~\eqref{mon1} and that
\begin{equation}
V = V(Y) = \frac{V_0}{2}(1-Y^2)^{-\frac{1}{1+\mu}}(2-\gamma Y^2),
\end{equation}
that $\tilde{Z}:= (1-Y^2)^{-\frac{1}{1+\mu}}(2-\gamma Y^2)(1-x^2)^{-1} >0$ on
${\bf S}$ satisfies $\tilde{Z}^\prime = - 6x^2\,\tilde{Z}$ and hence that
\begin{equation}\label{barZ}
\bar{Z}^\prime = - 6x^2\,\bar{Z}(1 - \bar{Z}), \qquad \bar{Z}:= \frac{\tilde{Z}}{1 + \tilde{Z}}
= \frac{2-\gamma Y^2}{(1-Y^2)^{\frac{1}{1+\mu}}(1-x^2) + 2 - \gamma Y^2}.
\end{equation}
Thus $\bar{Z}$ is a bounded monotone decreasing function on ${\bf S}\cap
\mathrm{dS}_0$ when $2 - \gamma(1+\mu) \geq 0$ ($\beta\geq 0$) and on ${\bf
S}\cap \mathrm{dS}_0\cap \mathrm{dS}_+\cap \mathrm{dS}_-$ when $2 -
\gamma(1+\mu) < 0$ ($\beta<0$), since the fixed points are the only invariant
subsets on $x=0$.

\begin{figure}[ht]
\centering
        \subfigure[\mbox{$\gamma = \frac43,\,\mu = \frac14$}.]{
        \label{period1a}
        \includegraphics[height=0.37\textwidth]{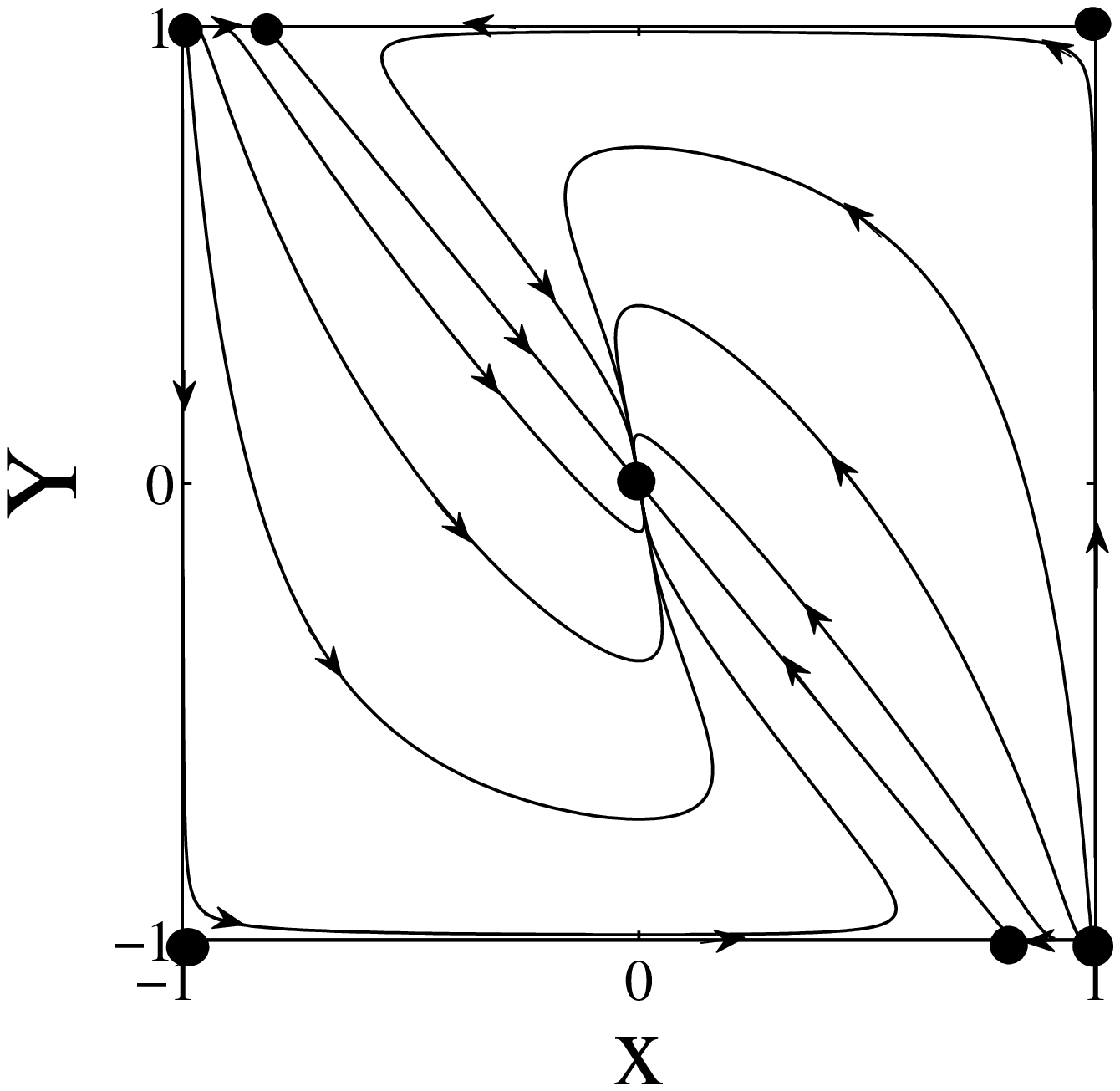}}\qquad
        \subfigure[\mbox{$\gamma = 1,\,\mu = 1$}.]{
        \label{period1b}
        \includegraphics[height=0.37\textwidth]{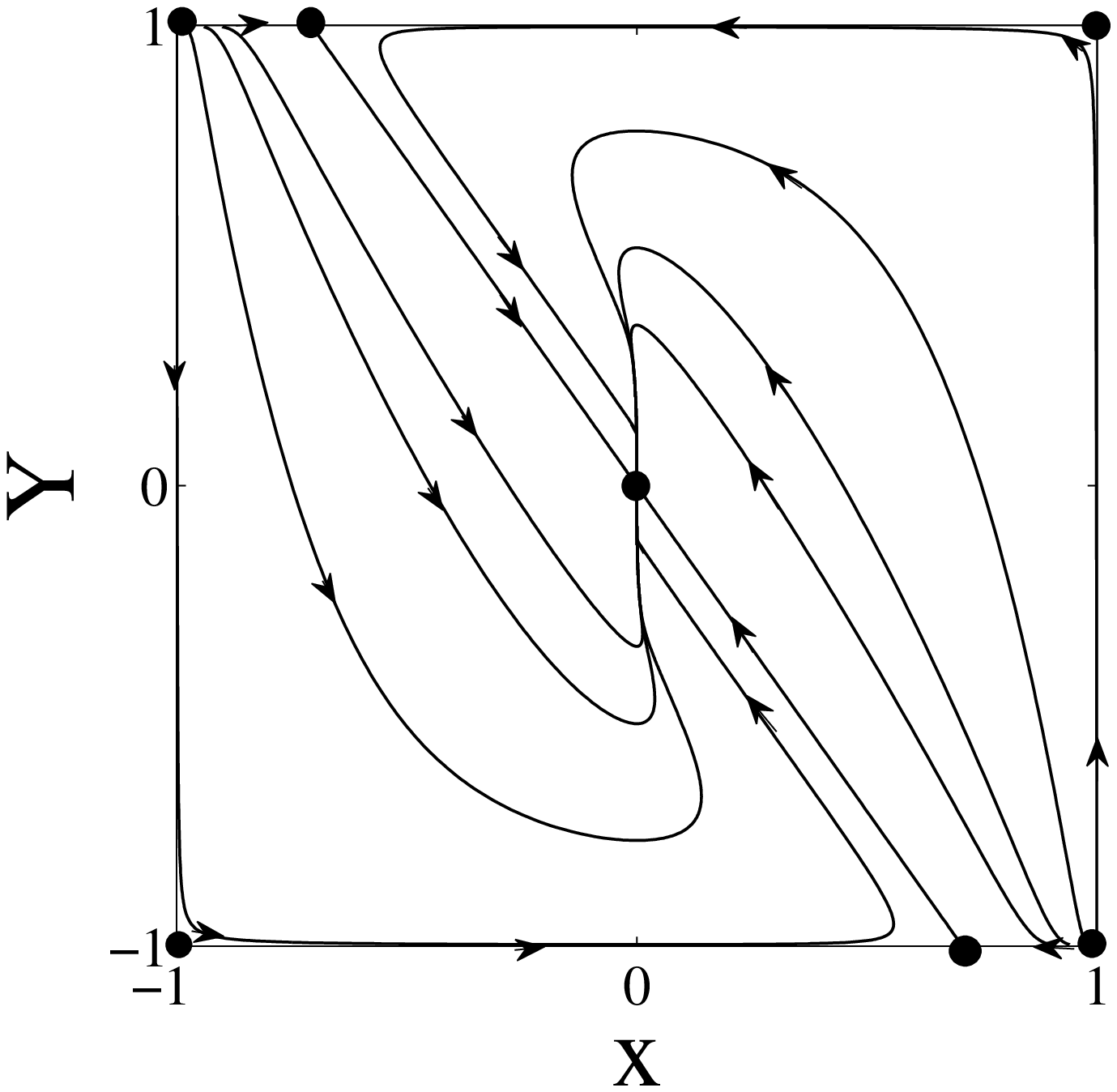}}\\
        \subfigure[\mbox{$\gamma = 1,\,\mu = 2$}.]{
        \label{period2}
        \includegraphics[height=0.37\textwidth]{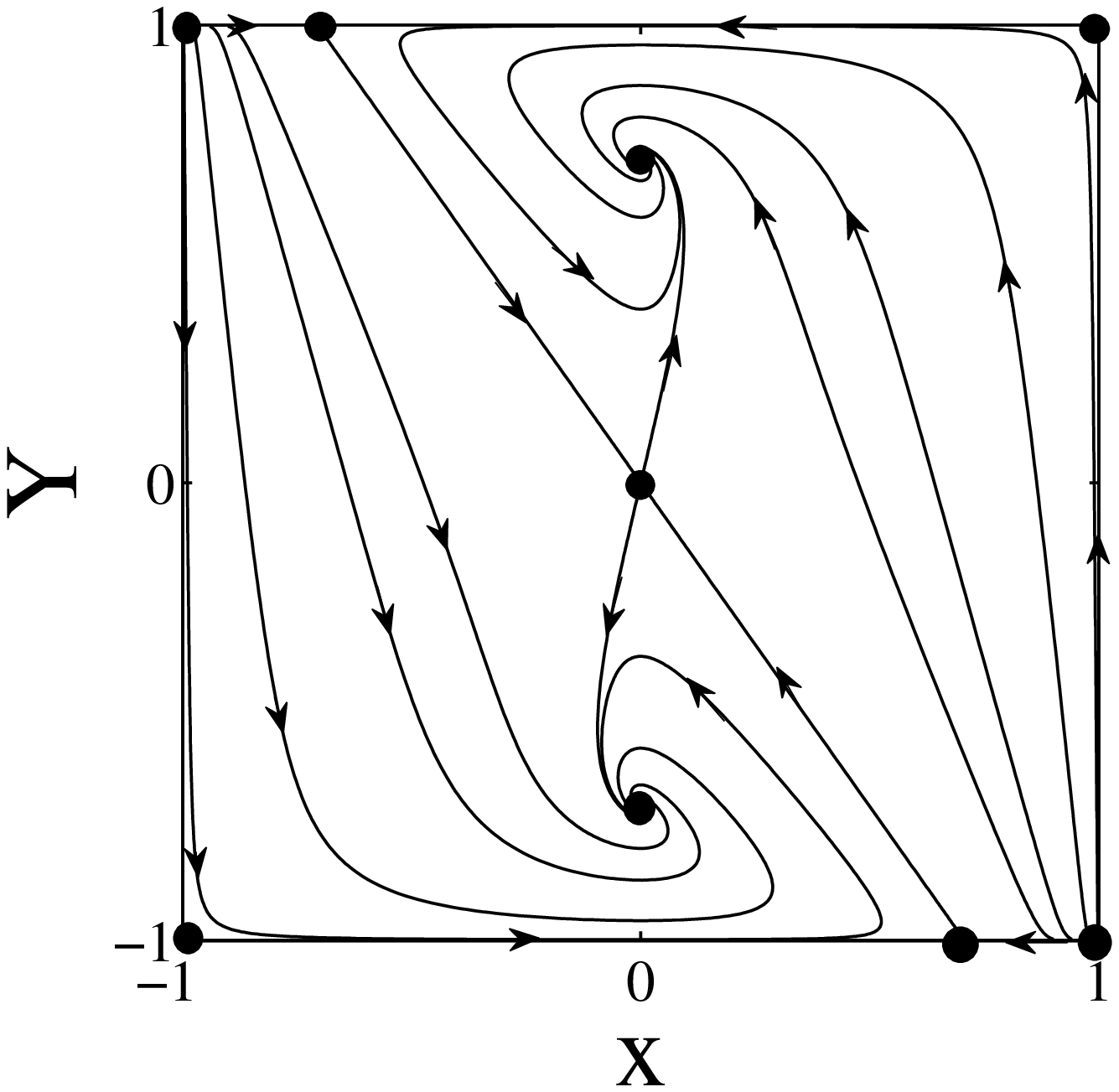}}
        \caption{Representative phase planes for the three cases characterized by
        $\beta>0$ ($\gamma = \frac43,\,\mu = \frac14$), $\beta=0$ ($\gamma = 1,\,\mu = 1$) and
        $\beta<0$ ($\gamma = 1,\,\mu = 2$), where $\beta= \frac{1}{1+\mu} - \frac{\gamma}{2}$.}
    \label{fig:phaseplanes}
\end{figure}

As a consequence, when $2 - \gamma(1+\mu) \geq 0$, i.e., when $\beta \geq 0$,
all solutions in ${\bf S}$, except for the fixed point $\mathrm{dS}_0$,
originate from the boundary $\partial{\bf S}$ where $\bar{Z}$ has its maximum
$\bar{Z} = 1$. It also follows that when $\beta<0$ all solutions also
originate from the boundary $\partial{\bf S}$, except for the fixed points
$\mathrm{dS}_*$ and the two solutions that originate from $\mathrm{dS}_0$.
Combining this with the structure on the boundary, and the hyperbolic
eigenvalues of the fixed points on $\partial{\bf S}$, it follows that when
$\beta \geq 0$ all orbits in ${\bf S}$ originate from the sources
$\mathrm{M}_{x=\mp 1}^{Y=\pm 1}$ except for the two perfect fluid
trajectories ${\cal PF}_\pm$ which originate from $\mathrm{PF}_\pm$.
Furthermore, the monotone function obtains its minimum $\bar{Z} = \frac23$ at
$\mathrm{dS}_0$ which is the future attractor for all orbits in ${\bf S}$,
i.e., they all have $\mathrm{dS}_0$ as their $\omega$ limit (i.e.,
$\mathrm{dS}_0$ is a global sink on ${\bf S}$, even though it is a
non-hyperbolic fixed point when $\beta=0$). Note that this is the case the
Chaplygin gas belongs to. On the other hand, when $\beta < 0$ all solutions
end at the hyperbolic sinks $\mathrm{dS}_\pm$ except for the perfect fluid
trajectories ${\cal PF}_\pm$, which end at $\mathrm{dS}_0$, as mentioned
above. Note also that the unstable manifold of $\mathrm{dS}_0$ yields two
solution trajectories that end at $\mathrm{dS}_\pm$, respectively, which
follows from the monotone function in combination with that $\mathrm{dS}_0$
is a hyperbolic saddle in this case. These results are illustrated by the
phase plane portraits given in Figure~\ref{fig:phaseplanes}.
\begin{remark}
The present statements concerning the use of the monotone function can be
formalized by means of the Monotonicity Principle~\cite{waiell97}, which is
stated as follows: Let $\phi_t$ be a flow on $\mathbb{R}^n$ with $S$ an
invariant set. Let $Z : S \rightarrow \mathbb{R}$ be a $C^1$ function whose
range is the interval $(a,b)$ where $a<b$. If $Z$ is decreasing on orbits in
$S$, then for all ${\bf x} \in S$ the $\omega$ and $\alpha$ limits belong to
the boundary of $S$ according to
\begin{align}
\omega({\bf x}) \subseteq \{ {\bf s} \in \bar{S}\!\setminus\! S|\lim_{{\bf
y}\rightarrow {\bf s}} Z({\bf y}) \neq b \},\nonumber \\
\alpha({\bf x}) \subseteq \{ {\bf s} \in \bar{S}\!\setminus\! S|\lim_{{\bf y}\rightarrow
{\bf s}} Z({\bf y}) \neq a \}.\nonumber
\end{align}
The existence of a monotone function on $S$ therefore e.g. excludes any
periodic orbits in $S$.
\end{remark}
%

\section{Discussion}\label{disc}

We have here shown that the perfect fluid solution that generates the scalar
field problem of the modified Chaplygin case appears as two straight lines in
the phase plane. Moreover, the two straight lines originate from fixed points
that correspond to the scale invariant perfect fluid solution with a linear
equation of state $p=(\gamma-1)\rho$ on 1-dimensional boundaries that
describe the scalar field problem corresponding to an exponential potential,
which this scale-invariant solution generates. \emph{All} other scalar field
solutions originate from fixed points that correspond to scale-invariant
massless scalar field solutions (except for the two solution that originate
from the de Sitter fixed point $\mathrm{dS}_0$ and ending at the de Sitter
fixed points $\mathrm{dS}_\pm$ when $\beta<0$). Hence \emph{no} other
solutions behave like the perfect fluid solutions towards the past limit
(which perhaps is a somewhat mote issue in the present context since the
purpose of these particular models is to describe intermediate and late stage
behavior).

The role of the perfect fluid solution in the scalar field case for the
Chaplygin gas has been previously discussed in~\cite{peretal04}
and~\cite{goretal05} and hence the role of the perfect fluid solution in the
case $\beta \geq 0$ deserves some comments ($\beta \geq 0$ is assumed
throughout the subsequent discussion). Let us first give the linearization of
${\cal PF}$:
\begin{subequations}
\begin{align}
(\sqrt{2}\,x + \sqrt{\gamma}\,Y)^{-1}(\sqrt{2}\,x + \sqrt{\gamma}\,Y)^\prime|_{x =-\sqrt{\gamma/2}\,Y} &=
\frac32\left[2\gamma\,Y^2 - R\left(\frac{2+\gamma\,Y^2}{2 - \gamma\,Y^2}\right)\right],\label{pfstab}\\
(\sqrt{2}\,x + \sqrt{\gamma}\,Y)^{-1}(\sqrt{2}\,x + \sqrt{\gamma}\,Y)^\prime|_{\mathrm{PF}_\pm} &= - \sfrac32(2-\gamma),\\
(\sqrt{2}\,x + \sqrt{\gamma}\,Y)^{-1}(\sqrt{2}\,x + \sqrt{\gamma}\,Y)^\prime|_{\mathrm{dS}_0} &= -\sfrac32\left[2 - \gamma(1+\mu)\right],
\end{align}
\end{subequations}
where $R= 2 - \gamma - \gamma\mu(1-Y^2)$. Here the values at the fixed points
$\mathrm{PF}_\pm$ and ${\mathrm{dS}_0}$ are just one of the eigenvalues at
each fixed point and show that the perfect fluid submanifold ${\cal PF}$ is
stable at $\mathrm{PF}_\pm$ and at $\mathrm{dS}_0$ when $\beta>0$. Note,
however, that the sign of the right hand side of~\eqref{pfstab} depends on
the value of $Y^2$ (and the values of $\mu$ and $\gamma$), as illustrated by
the Chaplygin case $(\sqrt{2}\,x + Y)^{-1}(\sqrt{2}\,x + Y)^\prime|_{x
=-\sqrt{1/2}\,Y} = \frac32\,\frac{Y^2(3Y^2-2)}{2 - Y^2}$. As a consequence,
solutions nearby $\mathrm{PF}_\pm$ drift away slightly from ${\cal PF}_\pm$
during part of their intermediate evolution, as seen in
Figure~\ref{fig:phaseplanes}. This means that the perfect fluid submanifold
${\cal PF}$ is not stable everywhere, which is a physical effect that can be
measured in terms of $x$ and hence the deceleration parameter $q$.
Nevertheless, thanks to that $\mathrm{dS}_0$ is a sink, even in the $\beta=0$
case, solutions that are close to the fixed points $\mathrm{PF}_\pm$ stay
close to the trajectories ${\cal PF}_\pm$ throughout their subsequent
evolution. Indeed, there is an open set of solutions that are arbitrarily
close to the perfect fluid solution throughout their intermediate and late
time evolution when $\beta \geq 0$, which can be seen as follows.

Consider the two finite heteroclinic chains that start from the source
$\mathrm{M}_{x=-1}^{Y=1}$ along the boundary $\partial{\bf S}$ that are
described by\footnote{A heteroclinic chain is a concatenation of heteroclinic
orbits (solution trajectories that begin and end at two distinct fixed
points), where the `final' ($\omega$ limit) fixed point of one solution
trajectory is the `initial' ($\alpha$ limit) fixed point of the next solution
trajectory.}
\begin{subequations}\label{chains}
\begin{align}
\mathrm{M}_{x=-1}^{Y=1} & \longrightarrow \mathrm{PF}_+ \longrightarrow \mathrm{dS}_0, \\
\mathrm{M}_{x=-1}^{Y=1} & \longrightarrow \mathrm{M}_{x=-1}^{Y=-1} \longrightarrow \mathrm{PF}_- \longrightarrow \mathrm{dS}_0,
\end{align}
\end{subequations}
and similarly for $\mathrm{M}_{x=1}^{Y=-1}$. As follows from the regularity
of the dynamical system, continuity, and the stability properties of the
fixed points, there exist two open sets of solutions that stay arbitrarily
close to the heteroclinic chains in~\eqref{chains} throughout their entire
history (and similarly for the analogous equivalent chains associated with
$\mathrm{M}_{x=1}^{Y=-1}$). As a consequence, these open sets of solutions
describe an open set of solutions that behaves like the perfect fluid
solution during the solutions \emph{intermediate and future evolution} when
$\beta\geq 0$, i.e., when $\mathrm{dS}_0$ is the future attractor. However,
there also exists an open set which only behave like the perfect fluid
solution toward their asymptotic future, described by $\mathrm{dS}_0$. In the
case of $\beta<0$, the chains continue with the heteroclinic orbits that go
from $\mathrm{dS}_0$ to $\mathrm{dS}_\pm$, and in this case there exists an
open set of solutions that behave like the perfect fluid solution at an
intermediate stage, but not toward the asymptotic future. On the other hand,
in this case there also exists an open set of solutions that \emph{never}
behave like the perfect fluid solution.

It should be stressed that the current results are \emph{not} in
contradiction to those in~\cite{goretal05}. There the authors introduced a
function involving the Chaplygin equation of state which was zero for the
Chaplygin case. By studying the evolution of this quantity they came to the
conclusion that the Chaplygin case was stable. The idea to study stability by
means of a function that reflects the scalar field generating solution's
equation of state characteristics is an interesting one, but the connection
with the stability of the corresponding solution in a state space picture is
not straightforward. Indeed, even to make a comparison demands that the
function is dimensionally compatible with the state space variables one uses
to describe the solution space. The present variables are dimensionless
(under conformal weight) while the function used in~\cite{goretal05} to
analyze the Chaplygin case carried dimension. To be able to make a comparison
with the present phase space picture and the type of analysis done
in~\cite{goretal05} therefore requires changing the `equation of state
function' to a dimensionless one. Such a dimensionless function was used
in~\cite{goretal05} in the case of a perfect fluid with linear equation of
state, which makes a comparison possible. The conclusion in~\cite{goretal05}
was that the perfect fluid solution is stable, and this precisely corresponds
to the local analysis of $\mathrm{PF}_\pm$ on the ${\cal E}_\pm$ subsets,
which indeed yields that the fixed points $\mathrm{PF}_\pm$ are stable
\emph{on} ${\cal E}_\pm$.


The presently studied example has shown a connection between the shape of the
potential for finite values of the field and bifurcations in the dynamical
systems picture. The same holds true when $\phi \rightarrow \pm \infty$. In
this case we have a bifurcation when $\gamma=2$. This corresponds to that the
fixed point sources are replaced by a heteroclinic cycle described by the
boundary as the $\alpha$-limit set. This is due to that the potential walls
become sufficiently steep so that oscillations take place towards the past.
This can be seen from considering the case of a scalar field with two
exponential terms with opposite signs, as done in~\cite{fos98}. This suggests
that scalar field problems can  be classified in terms of the properties of
the extremum properties of the potential for finite values of the field
(see~\cite{fos98b}) and the properties of the potential when the field goes
to infinity in terms of bifurcations in \emph{global regularized} dynamical
systems treatments, with subclassifications based on where and how the
bifurcations occur in the associated dynamical systems pictures (in a more
general context classifications involve several matter and geometrical
degrees of freedom). Due to the correspondence with some modified gravity
theories and scalar field problems in general relativity, similar
classifications are presumably possible for some gravity theories as well.
Note, however, that in the present global regularized dynamical systems
treatment, all bifurcations are associated with physical changes, such as
changing a minimum of the potential to a maximum. The world regularized,
including the sense that all fixed points are hyperbolic (or that fixed point
sets are transversally hyperbolic) to the extent this is physically possible,
is an essential demand for any sensible classification; classifications
involving non-hyperbolic fixed points that are consequences of `bad' choices
of variables, would be of little use.

The present example of the modified Chaplygin gas presumably illustrates some
quite general features as regards the relationship between solutions,
associated with some conditions, e.g. observational ones, and the field
theoretic descriptions they might generate. For example, as particular
solutions the `field generating solutions' may only describe part of the
temporal behavior of most solutions, and only for special initial data (and
sometimes even none of the temporal behavior for an open set of solutions, as
illustrated by the present $\beta<0$ case). It is only if the field
generating solutions originate at a source and end at a sink they might
describe global temporal behavior for an open set of solutions, and even then
in general only for, in some sense, special initial data.

\label{bibbegin}

\end{document}